\documentclass[sn-mathphys]{sn-jnl}% Math and Physical Sciences Reference Style
\jyear{2021}%

\theoremstyle{thmstyleone}%
\theoremstyle{thmstyletwo}%

\theoremstyle{thmstylethree}%

\usepackage{comment}
\usepackage{amsmath,amssymb,amsfonts}
\usepackage{graphicx}
\usepackage{textcomp}
\usepackage{xcolor}
\usepackage{hyperref}

\begin{document}
\title[Robust Quantum Arithmetic Operations with Intermediate Qutrits in the NISQ-era]{Robust Quantum Arithmetic Operations with Intermediate Qutrits in the NISQ-era}

\author*[1,2]{\fnm{Amit} \sur{Saha}}\email{abamitsaha@gmail.com}

\author[3]{\fnm{Anupam} \sur{Chattopadhyay}}\email{anupam@ntu.edu.sg}

\author[1]{\fnm{Amlan} \sur{Chakrabarti}}\email{acakc@caluniv.ac.in}

\affil*[1]{\orgdiv{A. K. Choudhury School of Information Technology}, \orgname{University of Calcutta}, \orgaddress{ \city{Kolkata}, \postcode{700106}, \state{West Bengal}, \country{India}}}

\affil[2]{ \orgname{Atos}, \orgaddress{\city{Pune}, \postcode{411057}, \state{Maharashtra},  \country{India}}}

\affil[3]{\orgdiv{School of Computer Science and Engineering}, \orgname{Nanyang Technological University}, \orgaddress{\city{Singapore},  \country{Singapore}}}


\abstract{Numerous scientific developments in this NISQ-era (Noisy Intermediate Scale Quantum) have raised the importance for quantum algorithms relative to their conventional counterparts due to its asymptotic advantage. For resource estimates in several quantum algorithms, arithmetic operations are crucial. With resources reported as a number of Toffoli gates or T gates with/without ancilla, several efficient implementations of arithmetic operations, such as addition/subtraction, multiplication/division, square root, etc., have been accomplished in binary quantum systems. More recently, it has been shown that intermediate qutrits may be employed in the ancilla-free frontier zone, enabling us to function effectively there. In order to achieve efficient implementation of all the above-mentioned quantum arithmetic operations with regard to gate count and circuit-depth without T gate and ancilla, we have included an intermediate qutrit method in this paper. Future research aiming at reducing costs while taking into account arithmetic operations for computing tasks might be guided by our resource estimations using intermediate qutrits. Therefore, the enhancements are examined in relation to the fundamental arithmetic circuits. The intermediate qutrit approach necessitates access to higher energy levels, making the design susceptible to errors. We nevertheless demonstrate that the percentage decrease in the probability of error is significant due to the fact that we achieve circuit efficiency by reducing circuit-depth in comparison to qubit-only works.}




\maketitle

\section{Introduction}
With an asymptotic advantage, quantum computers are predicted to outperform conventional computers in some computational tasks \cite{chuang}. Albeit, in this NISQ-era \cite{Preskill_2018}, quantum computers are not scalable enough, hence, we need to restrict ourselves to ancilla-free frontier zone \cite{gokhalefirst}. Therefore, despite the favourable asymptotic improvement that quantum algorithms for computationally challenging tasks have, precise run-time estimates are frequently lacking as a result of the scarcity of effective implementations of the quantum arithmetic operations that are used as functions in quantum algorithms \cite{hner2018optimizing, Chakrabarti_2021, amitderivative}.  By incorporating higher energy levels of quantum state into our advanced arithmetic circuit designs, we have attempted to overcome these problems.

The purpose of this study is to decompose arithmetic operations and provide the groundwork for more effective arithmetic algorithms. These arithmetic operations consist of multiple number of Toffoli gates. Decomposition of Toffoli gate is essential to implement these arithmetic operations. Thus it is of utmost importance to figure out which Toffoli decomposition technique out of many existing ones \cite{Selinger_2013, amy, Gokhale_2019} works best in terms of circuit fidelity in this NISQ-era. By breaking down Toffoli gates with the use of an intermediate qutrit approach \cite{Gokhale_2019, adderqudit}, the approaches suggested in this research do aim to carry out an asymptotic reduction of the complexity suggested in the study and enhance the error limits of the algorithm in its cutting-edge form. Qutrit systems \cite{Wang_2020} have already been successfully realized on superconducting \cite{PhysRevA.76.042319}, trapped ion \cite{qutrit} and photonic systems \cite{Gao_2020}, proving that quantum systems other than qubit-only systems may be thought of at higher levels. 

Therefore, we are bringing attention to the efficiency and error tolerance improvement while shifting to higher dimensional quantum systems. Since the Toffoli gate decomposition method requires access to a higher energy level, the design is prone to errors. Yet, by decreasing gate count and circuit depth compared to earlier studies, we show that there is significant overall reduction in the likelihood of error.

In brief, these are our specific contributions:
\begin{itemize}
    \item We deploy intermediate higher-dimensional quantum systems (qutrits) to accomplish fundamental quantum arithmetic operations, and we estimate the necessary resources to demonstrate that, in terms of circuit depth and circuit reliability, our method is superior to the existing approaches.
    \item By getting a percentage improvement in the probability of success, our numerical study proves that the deployment of intermediate qutrit for quantum arithmetic operations is sublimer to the works already in existence.
    %by $\simeq 40\%$ for Toffoli count 30 in a circuit.
\end{itemize}
The remainder of this paper is structured as follows: The required introductions to quantum arithmetic operations are presented in Section \ref{arith}. Section \ref{intermediate} analyses the resources needed for suggested quantum arithmetic operations with intermediate qutrit approach. Section \ref{error} presents the performance of the quantum arithmetic operations with intermediate qutrits under various forms of noise. Our last thoughts are summarized in Section \ref{conclusion}.
\section{Quantum Arithmetic Operations}\label{arith}
The history of basic quantum arithmetic operations is highlighted in this section.

\begin{figure}[h!]
    \centering
    \includegraphics[width=0.4\textwidth]{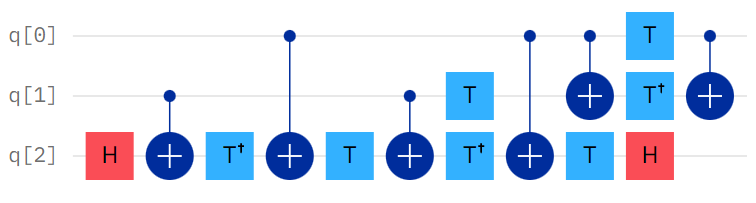}
    \caption{Decomposed Toffoli (Circuit Depth 12, T-count 7, CNOT count 6).}
    \label{fig:tof_selinger}
\end{figure}

\begin{figure}[h!]
    \centering
    \includegraphics[width=0.4\textwidth]{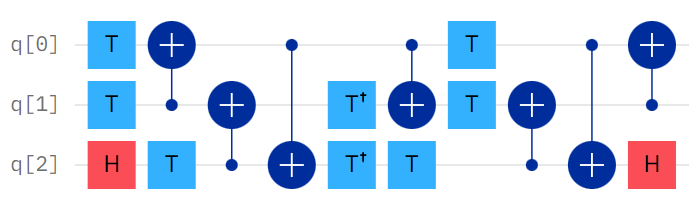}
    \caption{Decomposed Toffoli (Circuit Depth 10, T-count 7, CNOT-count 7).}
    \label{fig:tof_selinger1}
\end{figure}

\begin{figure}[h!]
    \centering
    \includegraphics[width=0.4\textwidth]{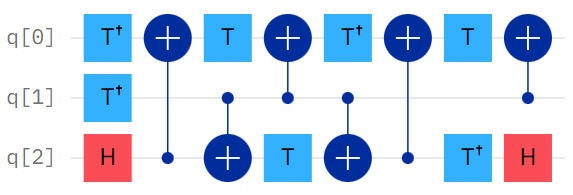}
    \caption{Decomposed Toffoli (Circuit Depth 8, T-count 7, CNOT-count 6).}
    \label{fig:tof_selinger2}
\end{figure}

\begin{figure}[h!]
    \centering
    \includegraphics[scale=.3]{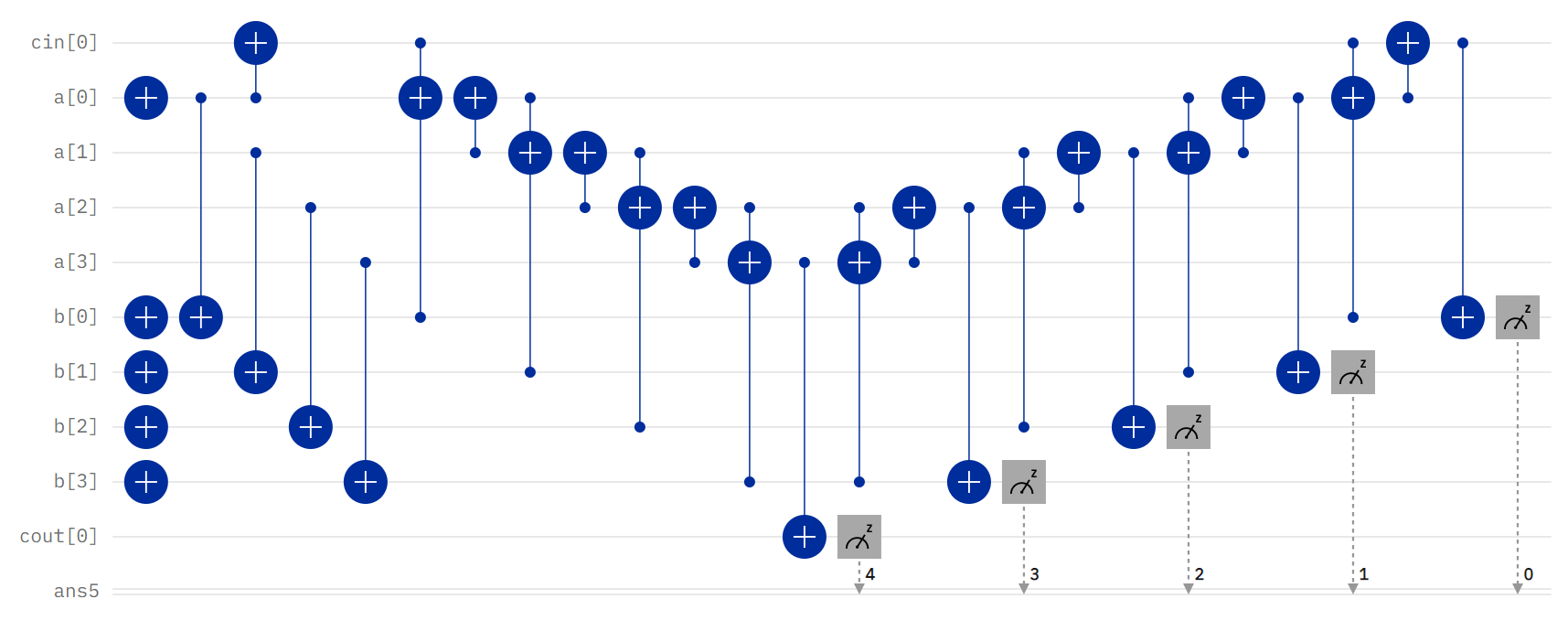}
    \caption{Circuit for 10-qubit Adder.}
    \label{adder}
\end{figure}

\begin{figure*}[h!]
    \centering
    \includegraphics[scale=.115]{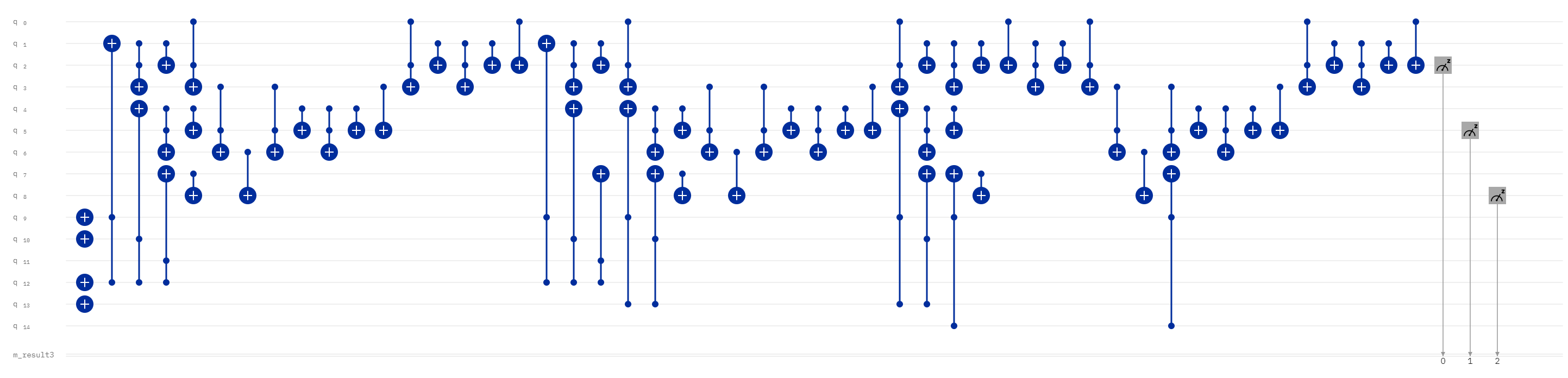}
    \caption{Circuit for 15-qubit Multiplier.}
    \label{multi}
\end{figure*}

\begin{figure}[h!]
    \centering
    \includegraphics[scale=.38]{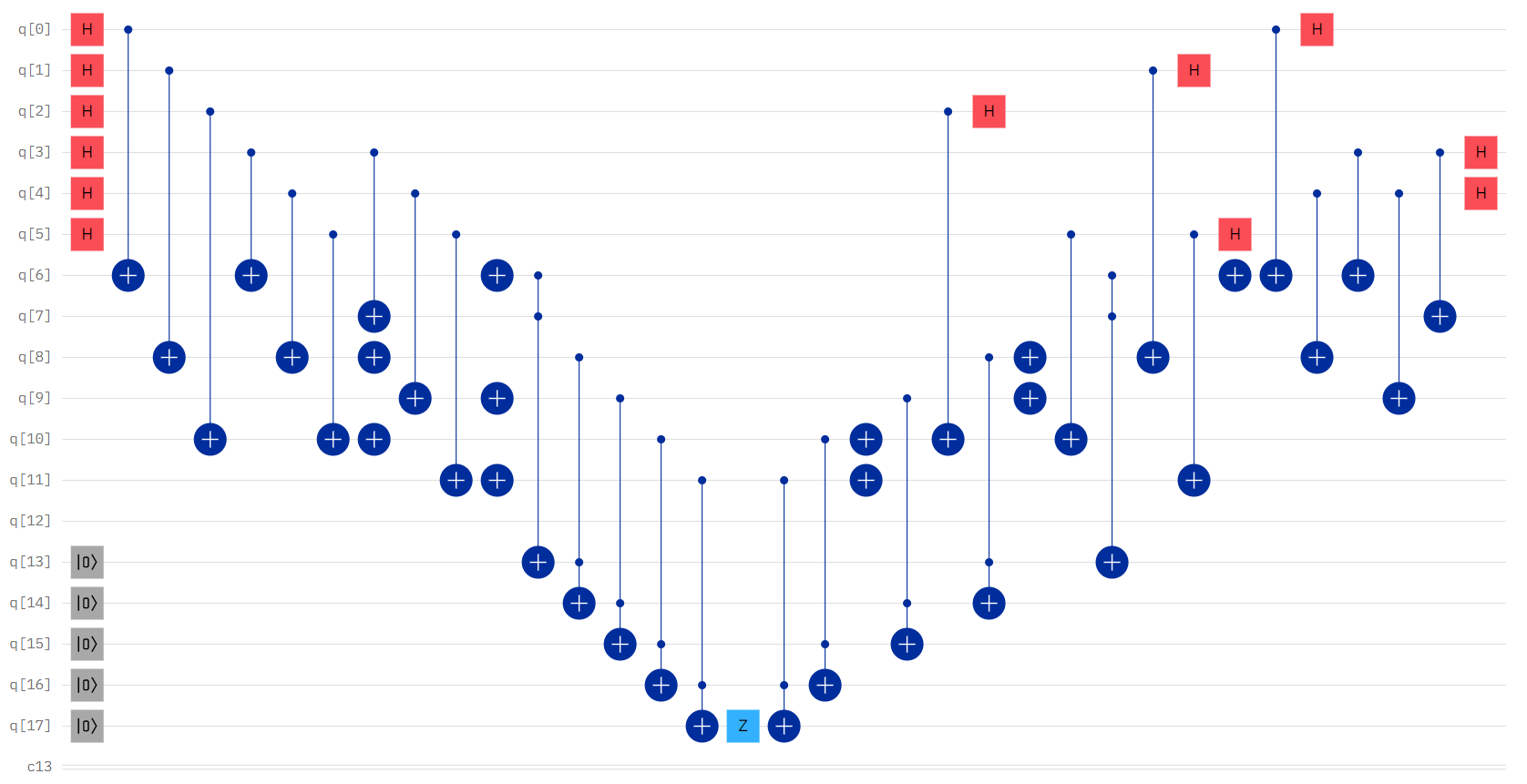}
    \caption{A part of circuit for 18-qubit Square root.}
    \label{sqrt}
\end{figure}

In different quantum algorithms, these operations are employed for resource estimates and error analysis.
Resources are often given as Toffoli costs in this section. As per the state-of-the-art works on quantum arithmetic operations, we need to decompose the Toffoli gate in a Clifford + T gate set \cite{Selinger_2013, amy} to estimate the resources.  Since this article deals with ancilla-free circuit implementation, we consider the decomposition of the Toffoli gate without ancilla. The three mostly used Toffoli decomposition with Clifford + T gate set and without ancilla are portrayed in Figures \ref{fig:tof_selinger}, \ref{fig:tof_selinger1} and \ref{fig:tof_selinger2}. As shown in Figures \ref{fig:tof_selinger}, \ref{fig:tof_selinger1} and \ref{fig:tof_selinger2}, the Toffoli decomposition in Figure \ref{fig:tof_selinger2} has circuit depth 8 (The number of gates required that has to execute in a path, is the length of that path. Thus, this path length is an integer. The longest path in the circuit is said to be the circuit depth), which is minimum as compared to the circuit depth of Figures \ref{fig:tof_selinger} and \ref{fig:tof_selinger1}, which are 12, 10. The T-count (the total number T gates) and CNOT-count (the total number of CNOT gates) of Figure \ref{fig:tof_selinger2} are in sync with  Figures \ref{fig:tof_selinger} and \ref{fig:tof_selinger1}.  This leads to the choice of Toffoli decomposition of Figure \ref{fig:tof_selinger2} as it has minimum circuit depth, which makes the circuit more noise-resilient in this NISQ-era. 
In this article, We perform all the calculations for quantum arithmetic operations in fixed-point arithmetic similarly to
\cite{hner2018optimizing}, which allows us to use the described quantum techniques for reversible function evaluation.
An $n$-bit representation of a number $x$ is
\begin{equation}
\label{eqn:fixed_point_repr}
x=\underbrace{x_{n-1}\cdots x_{n-p}}_p.\underbrace{x_{n-p-1} \cdots x_0}_{n-p},
\end{equation}
where $x_i \in \{0,1\}$ denotes the $i$-th bit of the binary representation of
$x$ and $p$ denotes the number of bits to the left of the binary decimal point. Let $Toffoli\_count$ denote the number of Toffoli gates required to compute an arithmetic function as per the state-of-the-art works. To perform quantum arithmetic operation, apart from Toffoli gates we may require some one or/and two-qubit gates (NOT or CNOT gate), but we concentrate on Toffoli gates only as the decomposition of the Toffoli gate contributes significantly to the depth of the quantum arithmetic circuit. Hence our focus is completely on the resources of the Toffoli decomposition, since circuit depth and gate costs are the pivot elements for error analysis of a circuit. Thus we need to derive the $Total\_depth$, $CNOT\_count$, $T\_count$ and $H\_count$ as the total circuit depth, the total CNOT gate count, the total T gate count and the total Hadamard count respectively from Toffoli count to analysis the probability of error in subsequent sections. As per the literature, we can try to parallelize the resulting circuits wherever possible to obtain the optimized depth with the help of ancilla qubits. Since we are dealing with NISQ-devices in this article, parallelization of the circuit is not feasible due to the limited number of available qubits. Hence, we also consider the number of ones in the binary expansion of $n$ is maximum $i.e.,$ $n$ number of ones, in other words, according to the existing works, the maximum resources that are required to implement the basic quantum operations are as follows:
\paragraph*{Addition/Subtraction}
As per \cite{Chakrabarti_2021, draper2006logarithmic}, one can perform addition of two
$n$-qubit registers in place with a Toffoli cost, 
\begin{equation}
%\resizebox{0.91\hsize}{!}{%
 Toffoli\_count_{add} =  4n-3\log_2n - 3\log_2(n-1) -10.
 %}%
\end{equation}
From the Toffoli count we derive as follows,
\begin{equation}
    %\resizebox{0.91\hsize}{!}{%
Total\_depth_{add}=32n-24\log_2n - 24\log_2(n-1) -80.
%$}% 
\end{equation}
\begin{equation}
 %   \resizebox{0.91\hsize}{!}{%
CNOT\_count_{add} =  24n-18\log_2n - 18\log_2(n-1) -60.
%$}%
\end{equation}
\begin{equation}
%\resizebox{0.91\hsize}{!}{%
 T\_count_{add} =  28n-21\log_2n - 21\log_2(n-1) -70.
 %$}%
\end{equation}
\begin{equation}
%\resizebox{0.91\hsize}{!}{%
 H\_count_{add} =  8n-6\log_2n - 6\log_2(n-1) -20.
%$}%
\end{equation}
For subtraction, the Toffoli count remains same as addition. An example of 10-qubit adder \cite{qasmbench} is portrayed in Figure \ref{adder}.
\paragraph*{Multiplication/Division} For multiplication we follow the method, which uses the controlled addition circuit in \cite{takahashi2009quantum} and requires a Toffoli count,
\begin{equation}
Toffoli\_count_{\text{mul}} = \frac{3}{2}n^2 + \frac{9}{2}n.
\end{equation}
\begin{equation}
Total\_depth_{\text{mul}} = 12n^2 + 36n.
\end{equation}
\begin{equation}
CNOT\_count_{\text{mul}} = 9n^2 + 27n.
\end{equation}
\begin{equation}
T\_count_{\text{mul}} = \frac{21}{2}n^2 + \frac{63}{2}n.
\end{equation}
\begin{equation}
H\_count_{\text{mul}} =3n^2 + 9n.
\end{equation}
This method can also be used for division, where the Toffoli count remains same as multiplication. An example circuit of multiplier from \cite{qasmbench} is portrayed in Figure \ref{multi}.
\begin{comment}
The fixed-point multiplication method's controlled additions necessitate ancilla qubits proportional to the register size, but the circuits support uncomputing the ancillas, allowing them to be reused for each consecutive addition that is not done in parallel. We may also parallelize each multiplication circuit by viewing each factor's register as $z \geq 1$ separate registers of size $n/z$, and each controlled addition for the $z$ subregisters can happen in parallel.
To aggregate the $z$ sub-results into the final result, $n \cdot (z-1)$ more qubits and $z-1$ additions are required.
$z=1$ indicates that no additional parallelization is used.
We can get a total T-depth cost of parallelized fixed-point multiplication by parallelizing the pairwise accumulation adds as well,
\begin{multline}
\label{eqn:t_depth_parallel_mult}
T\_depth_{\text{mul}}(n, z) = \lceil\frac{n}{z}\rceil \cdot (T\_depth_{add} + 6) + \lceil\log_2z\rceil \\ \cdot T\_depth_{add}.
\end{multline}
$(\text{T}_{\text{add}} + 6)$ is the T-depth of a controlled addition discussed in the Addition/Subtraction section.
\end{comment}
\paragraph*{Square Root} We employ the square root algorithm from \cite{hner2018optimizing}, which has the Toffoli count,
\begin{equation}
Toffoli\_count_{\text{sq}} = \frac{n^2}{2} + 3n - 4.
\end{equation}
\begin{equation}
    Total\_depth_{\text{sq}}(n)= 4n^2 + 24n - 32.
\end{equation} 
%where $2n+1$ qubits are required.
\begin{equation}
CNOT\_count_{\text{sq}} = 3n^2 + 18n - 24.
\end{equation}
\begin{equation}
T\_count_{\text{sq}} = \frac{7n^2}{2} + 21n - 28.
\end{equation}
\begin{equation}
H\_count_{\text{sq}} = n^2 + 6n - 8.
\end{equation}
A part of 18-qubit example circuit of square root from \cite{qasmbench} is portrayed in Figure \ref{sqrt}.
\section{Quantum Arithmetic Operations with Intermediate Qutrits}\label{intermediate}
\begin{figure}[h!]
    \centering
    \includegraphics[scale=.5]{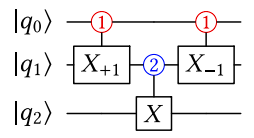}
    \caption{An example of Toffoli decomposition with intermediate qutrit, where input and output are qubits \cite{Gokhale_2019}.}
    \label{tof_qutrit}
\end{figure}

\begin{figure}[h]
    \centering
    \includegraphics[scale=0.26]{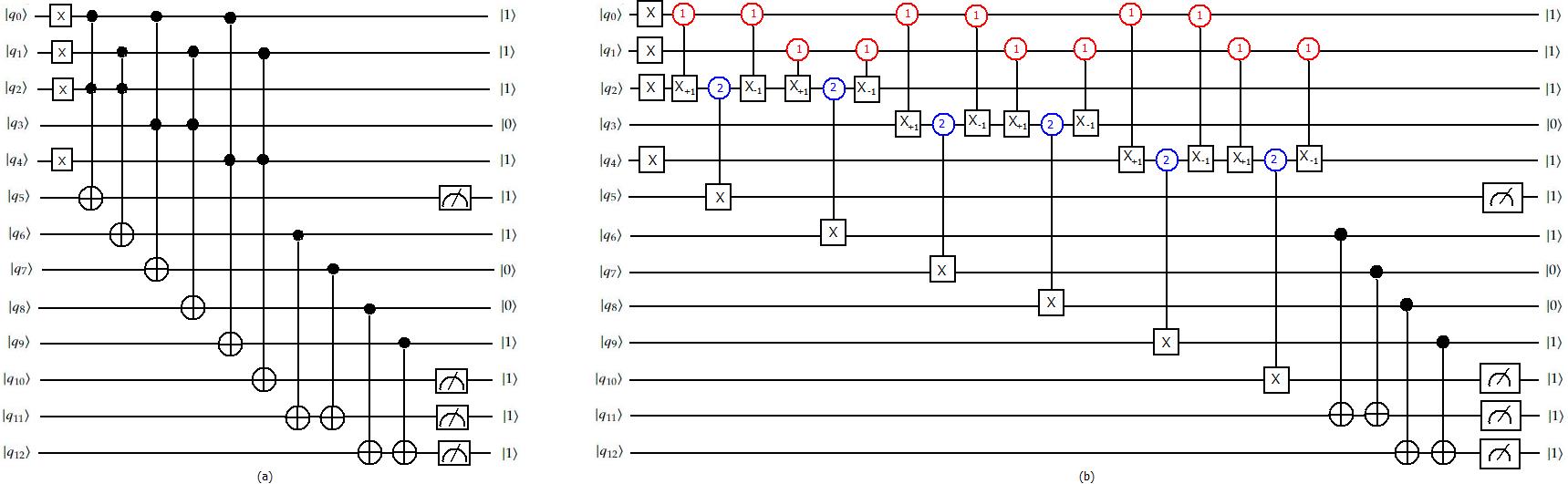}
    \caption{(a) Quantum Multiplier Circuit for the multiplication of $5 \times 3$; (b) Quantum Multiplier with Intermediate Qutrit for the multiplication of $5 \times 3$ \cite{amitderivative}.}
    \label{fig:multiplier}
\end{figure}

This section first describes the Toffoli decomposition with the help of intermediate qutrit that we have adapted from \cite{Gokhale_2019}. Thenceforth, we vividly discuss about quantum adder, multiplier and square root and their implementation with intermediate qutrits further in this section. In \cite{Gokhale_2019, 10.1145/3406309}, the authors showed that we may occupy the $\lvert2\rangle$ state temporarily during computation, hence temporarily ternary. Maintaining binary input and output allows this circuit construction to be inserted into any pre-existing qubit-only circuits.  A Toffoli decomposition via qutrits has been portrayed in Figure \ref{tof_qutrit} \cite{Gokhale_2019, 10.1145/3406309}. More specifically the goal is to carry out a NOT operation on the target qubit (third qubit) as long as the two control qubits, are both $\lvert1\rangle$. First, a $\lvert1\rangle$-controlled $X_{+1}$, where $+1$ is used to denote that the target qubit is incremented by $1 \ (\text{mod } 3)$, is performed on the first and the second qubits. This upgrades the second qubit to $\lvert2\rangle$ if and only if the first and the second qubits were both $\lvert1\rangle$. Then, a $\lvert2\rangle$-controlled $X$ gate is applied to the target qubit. Therefore, $X$ is executed only when both the first and the second qubits were $\lvert1\rangle$, as expected. The controls are reinstated to their original states by a $\lvert1\rangle$-controlled $X_{-1}$ gate, which reverses the effect of the first gate. The main result of this decomposition is that the temporary information may be stored in the $\lvert2\rangle$ state from ternary quantum systems instead of ancilla. Therefore, three generalised ternary CNOT gates with a circuit depth of three are sufficient to deconstruct the Toffoli gate. In actuality, no T gate is needed. 
As shown in the proposed approach, we have synthesized the quantum arithmetic operations via intermediate qutrits, in which, three generalized ternary CNOT gates are used to decompose the Toffoli gate without any ancilla. Hence, no T gate has been used in any proposed circuit synthesis. Therefore, T-depth becomes zero in our proposed approach. The overall circuit depth for Toffoli decomposition has now become three as compared to eight. Since there are no T gates in our proposed circuits, the resource analysis is based on the count of generalized ternary CNOT gate only, which is a Clifford gate. With this decomposition approach, we observe that the ternary CNOT count and the total circuit depth are same for Toffoli decomposition for quantum arithmetic operations. Let denote the ternary CNOT count as $Ternary\_CNOT\_count$ and analysis of the maximum resources required for quantum arithmetic operations with intermediate qutrits is as follows:
\paragraph*{Addition/Subtraction}
As per our approach, one can perform addition of two
$n$-qubit registers in place with a CNOT-cost, 
\vspace{-1mm}
\begin{equation}
%\resizebox{0.91\hsize}{!}{%
Ternary\_CNOT\_count_{add} =  12n-9\log_2n \\ - 9\log_2(n-1) -30.
%$}%
\end{equation}
For subtraction, the CNOT count remains same as addition.
\paragraph*{Multiplication/Division} For multiplication if one can follow our proposed method, then the CNOT count becomes,
\begin{equation}
Ternary\_CNOT\_count_{\text{mul}} = \frac{9}{2}n^2 + \frac{27}{2}n.
\end{equation}
This method can also be used for division, where the CNOT count remains same as multiplication. An example circuit of quantum multiplier with proposed decomposition is portrayed in Figure \ref{fig:multiplier} for better understanding.

\paragraph*{Square Root} If one can employ our proposed approach for the square root algorithm, then the CNOT count becomes,
\begin{equation}
Ternary\_CNOT\_count_{\text{sq}} = \frac{3n^2}{2} + 9n - 12.
\end{equation}
Next, we address the action of various error models on our decomposition of Toffoli gate for quantum arithmetic quantum operations.
\section{Error Analysis of Quantum Arithmetic Operations  with Intermediate Qutrits}\label{error}
Any finite-dimensional quantum system is susceptible to decoherence, noisy gates, and other types of errors. It has been demonstrated that switching from binary to higher dimensional states increases the system's error rate. In this section, we study how these errors affect the decomposition of the Toffoli gate for basic arithmetic operations. Qutrits cause an increase in error, although the total error probability of the decomposition is lower than that of the prior decomposition \cite{amy} because there are fewer gates, and a shallower structure.
\subsection{Generic Error Model}
The typical quantum error is for gate and relaxation error \cite{Gokhale_2019}, which can be realized by the Kraus Operator formalism \cite{chuang}. If the density matrix representation of a pure quantum state is $\sigma = \lvert{\psi}\rangle{\langle{\psi}\rvert}$, the evolution of this quantum state for any channel is represented as the function $\mathcal{E}(\sigma)$:
\vspace{-2mm}
\begin{align}
    \mathcal{E}\left(\sigma\right) = \mathcal{E}\left(\lvert{\psi}\rangle\langle{\psi}\rvert\right) = \sum_i K_i \sigma K_i^\dagger,
\end{align}
where $K_i$ are called the Kraus Operators, and $K_i^\dagger$ is the matrix conjugate-transpose of $K_i$, $\forall$ $i$. The development of a state under a noise model may also be represented using the Kraus operator formulation. Now we thoroughly discuss about the gate errors followed by relaxation error or idle error of a quantum system.
\subsubsection{Gate errors}
There are four potential error channels for a one-qubit gate in a binary quantum system with just one-qubit and two-qubit gates, which may be written as products of the two Pauli matrices, a NOT gate, $X = \begin{pmatrix}
0 & 1 \\
1 & 0
\end{pmatrix}$ and a phase gate, $Z = \begin{pmatrix}
1 & 0 \\
0 & -1
\end{pmatrix}$. The possible error channels are: (i) no-error $X^0 Z^0 = I$, (ii) the phase flip which is the product $X^0 Z^1$, (iii) the bit flip which is $X^1 Z^0$ and (iv) the phase+bit flip channel given by $X^1 Z^1$. The following is how we may formulate this one-qubit gate error model using the Kraus operator formalism:
\begin{align}
    \mathcal{E}(\sigma) = \sum_{j=0}^{1} \sum_{k=0}^{1} p_{jk} (X^j Z^k) \sigma (X^j Z^k)^{\dagger},
\end{align}
where $p_{jk}$ is the probability of the corresponding Kraus operator. 
%Let us assume all error terms have equal probabilities, i.e. $p_{jk} = p$ for $j,k \neq 0$. This is the standard symmetric depolarization noise model. Henceforth, we use $p_1$ and $p_2$ to indicate the probability of error for single and two qudit gates respectively.
An undesirable Pauli operator is followed by a noisy gate that is modelled as an ideal gate \cite{fowler2012surface}. In alternative words, an one-qubit gate is followed by an unwanted Pauli $\in \{X, Y, Z\}$ with probability $p_x, p_y, p_z$ respectively; and a two-qubit gate is followed by an unwanted Pauli $\in \{I, X, Y, Z\}^{\otimes 2} \setminus \{I,I\}$ with probability $p_i \cdot p_j$, where $i,j \in \{x, y, z\}$. We provide the one-qubit and two-qubit gate error probability as $p_1$ and $p_2$ respectively for convenience.
Unwanted $X$ and $Z$ Pauli errors, each of which comes in two varieties, can follow a one-qubit gate in a binary system. Therefore, there are $2^2 -1$ possible ways for an error to follow an one-qubit gate (without taking the identity error into account). The development of the system under noisy one-qubit operations is shown in Eq.~\ref{eq:single_gate_err}, where $p_1$ is the probability of a one-qubit Pauli error.
\vspace{-3mm}
\begin{equation}
%\resizebox{0.97\hsize}{!}{%
    \mathcal{E}(\sigma) = (1-(2^2-1)p_1)\sigma + \sum_{jk \in \{0,1\}^2 \setminus 0*2} p_{jk} K_{jk} \sigma K_{jk}^{\dagger},
    %$}%
\label{eq:single_gate_err}
\end{equation}
where $K_{jk}$ represents the various Pauli operators.
Similar to one-qubit gates, two-qubit gates have the potential to result in an undesirable Pauli operator on each of the two qubits. Therefore a gate can be noisy in $2^4-1$ different ways (not including the identification operation on both qubits). The development of the system under noisy two-qubit operations is shown in Eq.~\ref{eq:two_gate_err} if $p_2$ is the probability of two-qubit gate mistakes.
\begin{comment}
\begin{multline}
%\resizebox{0.97\hsize}{!}{%
    \mathcal{E}(\sigma) = (1-(2^4-1)p_2)\sigma \\+ \sum_{jklm \in \{0,1\}^2 \setminus 0*2} p_{jklm} K_{jklm} \sigma K_{jklm}^{\dagger}
\end{multline}
\end{comment}
\vspace{-3mm}
\begin{equation}
%\resizebox{0.97\hsize}{!}{%
    \mathcal{E}(\sigma) = (1-(2^4-1)p_2)\sigma \\+ \sum_{jklm \in \{0,1\}^2 \setminus 0*2} p_{jklm} K_{jklm} \sigma K_{jklm}^{\dagger},
    %$}%
    \label{eq:two_gate_err}
\end{equation} where $p_{jklm} = p_{jk} \cdot p_{lm}$. The underlying depolarizing channel's symmetry or asymmetry has no bearing on the likelihood that the density matrix stays error-free. Instead, it is determined by the overall probability of error.
The decomposition we utilise in this instance only applies to ternary quantum systems with two-qutrit gates. In general, our technique employs up to three dimensions to decompose a Toffoli gate. Consequently, the error in our system scales as $\mathcal{O}(3^4)$ as shown in Eq.~\ref{eq:two qudit gate error} for a $3$-dimensional system.
% As $(d+2)$-ary quantum systems have been employed for $n$-qudit MCT decomposition, we have a similar form of the two-qudit gate error channel for $(d+2)$-ary quantum systems:
\vspace{-2mm}
\begin{equation}
%\resizebox{0.98\hsize}{!}{%
   \mathcal{E}(\sigma) = \{1-(3^4-1)p_2\}\sigma \\+ \sum_{\substack{jklm \in \\ \{0,1,2, \dots, 3\}^4 \setminus 0000}} p_{jklm} K_{jklm} \sigma K_{jklm}^{\dagger}.
   %$}%
\label{eq:two qudit gate error}
\end{equation}
It can be inferred that the decrease in the probability of no-error for two-qutrit gates due to the usage of higher dimensions for ternary systems is $1-81p_2$ as compared to $1-15p_2$ for binary systems. Now we shed some lights on the idle error or relaxation error.
\begin{comment}
\begin{table}[!htb]
    \centering
    \caption{Probability of success of two-qutrit gates due to the usage of higher dimensions}
    \resizebox{8.7cm}{!}{%
    \begin{tabular}{|c|c|c|}
    \hline
   Dimension $d$ & without our decomposition & with our decomposition  \\
    \hline
    2  & $1 - 15p_2$  & $1 - 81p_2$ \\
    \hline
    \end{tabular}}
    \label{tab:compgate}
\end{table}
\end{comment}
\subsubsection{Idle error}
Idle errors mostly concern the relaxation from higher to lower energy levels in quantum systems. An alternative name for this is amplitude damping. Qutrits are irreversibly transported to lower states via this noise channel. The sole amplitude damping channel for qubits is between $\lvert1\rangle$ and $\lvert0\rangle$; we refer to this damping probability as $\lambda_1$. The Kraus operators for amplitude damping in qubits are:
\begin{align} K_0 = \begin{pmatrix} 1 & 0 \\ 0 & \sqrt{1 - \lambda_1} \end{pmatrix} \text{\quad and \quad} K_1 = \begin{pmatrix} 0 & \sqrt{\lambda_1} \\ 0 & 0 \end{pmatrix}
\end{align}
 We also model damping from $\lvert2\rangle$ to $\lvert0\rangle$ for qutrits, which happens with probability $\lambda_2$. The Kraus operator for amplitude damping for qutrits may be described as follows:
%\begin{align}
$$K_0 = \begin{pmatrix} 1 & 0 & 0 \\ 0 & \sqrt{1-\lambda_1} & 0 \\ 0 & 0 & \sqrt{1 - \lambda_2}  \end{pmatrix}
\text{, }$$
%\end{align}
\begin{align}
K_1 = \begin{pmatrix} 0 & \sqrt{\lambda_1} & 0 \\ 0 & 0 & 0 \\ 0 & 0 & 0  \end{pmatrix} \text{\quad and \quad}
K_{2} = \begin{pmatrix} 0 & 0  & \sqrt{\lambda_{2}}\\ 0 & 0 & 0 \\ 0 & 0 & 0 \end{pmatrix}.
\end{align}
In each Kraus Operator $K_i$, the value of $\lambda_i \propto exp(-t/T_{1_i})$, where $t$ is the computation's run-time, and $T_{1_i}$ are the relaxation time. In certain higher end IBM Quantum Devices \cite{ibmquantum}, the value is $T_{1_1} \simeq 100 \mu s$ for qubit systems. Moreover, the value is $30 \mu s$ for qutrit ($T_{1_2}$) quantum devices \cite{https://doi.org/10.48550/arxiv.2203.07369}. However, the length of time is determined by the depth of the circuit. As a result, idle error is reduced by decreasing the circuit depth. Since depth has been optimized, our decomposition's decoherence is much smaller than it was with the prior decomposition.
\subsection{Experimental Analysis}\label{prob_success}
\subsubsection{Experimental setup}
 The binary circuits of the quantum arithmetic operations are taken from \cite{qasmbench} in the form of QASM (Quantum Assembly language) and implemented on IBMQ \cite{ibmquantum} to verify its functionality. After decomposition of Toffoli gates with intermediate qutrits, the binary-ternary circuits of the quantum arithmetic operations are simulated and verified on QuDiet platform \cite{qudiet}. Finally, analysis of success probability has been numerically simulated on Python 3.7, with processor Intel(R) Core(TM) i5-6300U CPU \@ 2.40 GHz 2.50 GHz, RAM 8.00 GB, and 64-bit windows operating system.
\subsubsection{Analysis of success probability}
\begin{figure}[h!]
    \centering
    \includegraphics[scale=0.4]{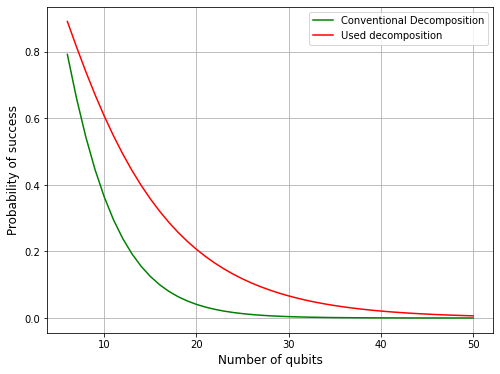}
    \caption{Probability of success for the decomposition of Toffoli gate using our used method versus the method in \cite{amy} for quantum addition/subtraction operation.}
    \label{fig:successadd}
\end{figure}

\begin{figure}[h!]
    \centering
    \includegraphics[scale=0.4]{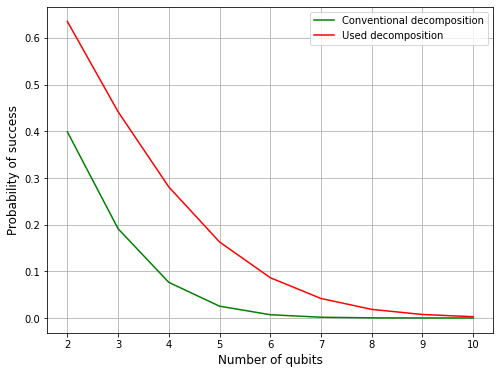}
    \caption{Probability of success for the decomposition of Toffoli gate using our used method versus the method in \cite{amy} for quantum multiplication/division operation.}
    \label{fig:successmul}
\end{figure}

\begin{figure}[h!]
    \centering
    \includegraphics[scale=0.4]{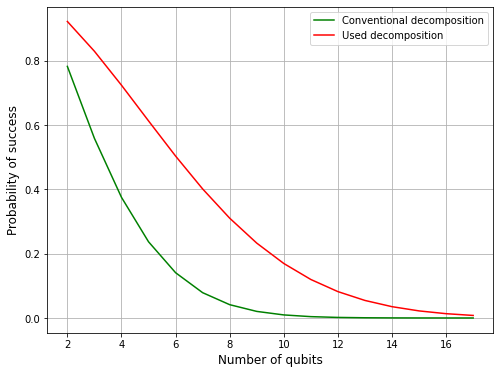}
    \caption{Probability of success for the decomposition of Toffoli gate using our used method versus the method in \cite{amy} for quantum square root operation.}
    \label{fig:successsqrt}
\end{figure}

The authors argued for the usage of higher dimension for the effective decomposition of Toffoli gates in \cite{Gokhale_2019}. They made the assumption that the value of $T_{1_i}$ for ternary systems is the same as for binary systems in that paper, although there was no qudit hardware available at the time. However, as stated in the article above, for ternary systems, we now have the value of $T_{1_3}$. The likelihood of success in the decomposed circuit of a Toffoli gate is therefore investigated in this section using both the approach in \cite{amy} and our way in this article. It's important to note that whereas the decomposition of \cite{amy} demands both one- and two-qubit gates, our technique just calls for two-qutrit ternary gates. Here, we have shown the differences between the approach in \cite{amy} and the one we used for the complexity of decomposing a Toffoli gate in terms of the number of gates and depth of the circuit. For each Toffoli decomposition as shown in Figure \ref{fig:tof_selinger2}, one requires 7 one-qubit gates and 6 two-qubit gates with overall circuit depth 8, whether we need 3 two-qutrit gates with circuit depth 3 for our used decomposition. 
As discussed in the preceding subsections, small errors in quantum circuit gates may be thought of as an ideal gate followed by an undesired Pauli operator. The likelihood that the circuit remains error-free (probability of success) as per \cite{majumdar2021optimizing} for the decomposition in \cite{amy} and our used decomposition is compared instead of the probability of small errors in the circuit, without sacrificing generality. For any decomposition, the generalized formula for probability of success ($P_{success}$) described in \cite{majumdar2021optimizing, amitpra} is the product that the individual components do not fail. Alternatively we can say that,
\begin{equation}
%\label{eq:success}
    P_{success} = \Pi_{gates} ({(P_{success~of~ gate})}^{\#~gates} \times  e^{-(depth/T_1)}), 
    \label{success}
\end{equation}
where the second term is the likelihood of no relaxation error, and the first term's product is the likelihood of using gates of all sorts (one-qubit, two-qubit, and two-qutrit) in the decomposition. When a certain kind of gate is not used in a decomposition, the related term has a value of 1 because of the zero in the power. For instance, the contribution of our proposed decomposition to the product is 1, as it does not require a one-qutrit gate.
The majority of current quantum devices are binary, and in the IBMQ Quantum Devices, the probability of one-qubit and two-qubit gates are in the range of $10^{-4}$ and $10^{-2}$ respectively \cite{ibmquantum}. Moreover, the time $T_{1_1}$  of the majority of IBM's quantum devices fall between $100 \mu s$.  However, in \cite{https://doi.org/10.48550/arxiv.2203.07369}, the authors experimentally exhibited that the value of $T_{1_2}$  for qutits is $30 \mu s$, and we consider the same for our estimation too. We assume that the probability of error of each two-qubit and two-qutrit gate is $10^{-2}$, that of one-qubit gate is $10^{-4}$, and the time $T_{1_1}$ is $100 \mu s$ and $T_{1_2}$ is $30 \mu s$ for our simulation.
%\begin{comment}
In Figures ~\ref{fig:successadd}, \ref{fig:successmul} and \ref{fig:successsqrt}, we exhibit the probability of success through numerical simulation with the help of Eq. \ref{success} for the Toffoli gate decomposition using the method of \cite{amy} (which we label as `conventional decomposition') and compare with it our used method for quantum arithmetic operations, addition/subtraction, multiplication/division and square root respectively. We find that our proposed technique produces much fewer errors than the decomposition in \cite{amy}. More specifically, there is an approximately 20\% decrease in error probability for adder with 20 number of qubits. Similarly for multiplier and square root, the in error probability is nearly 20\% for 5 number of qubits and 10 number of qubits respectively.
%\end{comment}
This is due to the fact that our decomposition has fewer gates and is shallower as compared to the conventional method. Although our decomposition employs a few qutrit gates, which have a higher error probability due to the curse of dimensionality, our technique is superior due to the overall large reduction in gate count and depth. As shown in Figures ~\ref{fig:successadd}, \ref{fig:successmul} and \ref{fig:successsqrt}, it can be inferred that for quantum arithmetic operations, our used decomposition has a probability of success higher than \cite{amy}. Thus, it can be concluded that the conventional decomposition methods attain more error, whereas our used decomposition method yields less erroneous results or higher fidelity for different quantum arithmetic operations. 

 Due to the unavailability of the error rate for the qutrit gates, we only consider decoherence time of a qutrit system to calculate the error probability in this paper. The results of this work are very intriguing to look out fault-tolerant arithmetic circuits as future research. It will be very interesting to investigate how QECC or fault-tolerance will come into play for qubit-qutrit systems in future.

\begin{comment}
A comparative study of our Toffoli decomposition used in derivative pricing with some previous works \cite{Chakrabarti_2021, Selinger_2013} is shown in Table \ref{tab:comp_resource}. Our work outperforms them in terms of T-depth, overall circuit depth, overall gate count of the circuit and probability of error.
\begin{table}[htb]
    \centering
    \caption{Comparative analysis between conventional approach \cite{Chakrabarti_2021, Selinger_2013} and our approach with qutrits used in derivative pricing problem for the decomposed circuit of Toffoli gate}
    \resizebox{8.5cm}{!}{%
    \begin{tabular}{|c|c|c|c|c|c|}
    \hline
         & \# T  & \# overall  & \# two-qutrit  & \# Gate  &  Prob. of error (\%) \\
         & depth & circuit depth & gates & count & for Toffoli count 30\\
         \hline
        Decomposition  & 1 & 7 & 0 & 25 & 99.95\\ 
        of \cite{Chakrabarti_2021, Selinger_2013} & & & & &\\
        \hline
        Our Decomposition & 0 & 3 & 3 & 3 & 60\\
        \hline
    \end{tabular}}
    \label{tab:comp_resource}
\end{table}
\end{comment}
\section{Conclusion}\label{conclusion}
By addressing fundamental quantum arithmetic operations and estimating the resources with intermediate higher dimensional qutrits, we showed that our methodology surpasses existing methods in terms of circuit depth and reliability. We have also investigated how various error models affect this decomposition method. Our research demonstrates that the few gates in higher dimensional quantum systems employed in the decomposition process are more error-prone. However, compared to traditional quantum arithmetic circuits, our enhanced gate count and circuit depth have allowed us to achieve low overall error probability, allowing the gates to function with excellent fidelity. Through numerical simulation, we have demonstrated, that in comparison to the traditional method of Toffoli decomposition on NISQ-devices, intermediate qutrit-based Toffoli decomposition acquired a percentage increase in the chance of success for quantum arithmetic operations. With the use of intermediate qudit and these enhanced quantum arithmetic operations, additional quantum circuits' decomposition may be studied in more depth in near future. 
\bibliography{biblio}% common bib file
%\bibliography{biblo}
\vspace{1cm}
\textbf{Statements and Declarations:} The authors declare that no funds, grants, or other support were received during the preparation of this manuscript. The authors have no relevant financial or non-financial interests to disclose. All authors contributed to the study conception and design. Material preparation, data collection and analysis were performed by Amit Saha, Anupam Chattopadhyay and Amlan Chakrabarti. The first draft of the manuscript was written by Amit Saha and all authors commented on previous versions of the manuscript. All authors read and approved the final manuscript. Our manuscript has no associated data.
\end{document}